# Fine Grain 3D Integration for Microarchitecture Design Through Cube Packing Exploration*

Yongxiang Liu[1], Yuchun Ma[2], Eren Kursun[1], Glenn Reinman[1] and Jason Cong[1,3]

[1]Computer Sience Department, University of California, Los Angeles
[2]Dept of Computer Science and Technology, Tsinghua University, P.R.China
[3]California NanoSystems Institute
E-mail: myc@mail.tsinghua.edu.cn; cong@cs.ucla.edu

**Abstract**

Most previous 3D IC research focused on "stacking" traditional 2D silicon layers, so the interconnect reduction is limited to inter-block delays. In this paper, we propose techniques that enable efficient exploration of the 3D design space where each logical block can span more than one silicon layers. Although further power and performance improvement is achievable through fine grain 3D integration, the necessary modeling and tool infrastructure has been mostly missing. We develop a cube packing engine which can simultaneously optimize physical and architectural design for effective utilization of 3D in terms of performance, area and temperature. Our experimental results using a design driver show 36% performance improvement (in BIPS) over 2D and 14% over 3D with single layer blocks. Additionally multi-layer blocks can provide up to 30% reduction in power dissipation compared to the single-layer alternatives. Peak temperature of the design is kept within limits as a result of thermal-aware floorplanning and thermal via insertion techniques.

## 1. Introduction

Microprocessor architecture design faces greater challenges in the deep submicron era due to interconnect and technology scaling issues. Each manufacturing process below 100nm has presented additional challenges, hence resulting in higher cost and process complications. As a result, three-dimensional integration technology gained significant interest in recent years due to the performance improvement it provides for existing process technologies.

An extensive amount of research has demonstrated the latency reduction through additional device layers [1, 2, 3, 4, 30], however most prior art on 3D [4, 5, 6] is restricted to stacking traditional 2D dies. Such stacking offers significant reduction in inter-block latency, whereas it does little to help *intra-block* wire latency. Although further improvement in both power and performance is possible from this type of fine-grain 3D integration, the lack of modeling and tool infrastructure has been the main reason for not utilizing these additional benefits.

Recent studies have provided block models for various architectural structures including 3D cache [9, 12, 13, 14], register file [10, 11] and instruction scheduler [30], along with 2D based placement tools such as [22]. However, these models are limited to folding blocks by wordlines or bitlines, and they do not explore further improvement from techniques such as port partitioning, which we shall discuss in Section 3. Other studies [4, 28, 22] simply extended the same 2D floorplan to multiple layers, which in turn causes heating problems due to vertical stacking of hotspots. None of the previous studies explore the overall performance impact of using combination of 2D and 3D implementations of different components in the same design.

Yet another critical issue that was not modeled in the prior art is thermal effects of multi-layer 3D stacking. Although multi-layer blocks can reduce the area, delay, and power consumption for each individual block [30, 22], the best configuration for each single block may not lead to the best design for the entire multi-layered chip.

In 3D design, the *co-optimization* of micro-architectural and physical design is absolutely essential, as most of the potential benefits are due to wirelength reduction. Recently, the MEVA-3D [4] framework was proposed to bridge the gap between physical planning and architectural design. This framework uses micro-architectural loop sensitivities in the floorplanning process to guide block placement. However, MEVA-3D is limited to single-layer blocks and does not enable multi-layer block exploration, where each block can span more than one silicon layers. We model the multi-layer 3D design as cube packing problem. Although some 2D packing representations have been extended to handle the cube packing problem, such as ST [7], 3D sub TCG [8], they merely optimize the volume of cubic blocks based on simulated annealing. However, the packing volume is not the major issue in architectural design, additional design optimization needs to target the intra-block wire latency and the corresponding performance improvement.

In this paper, we explore finer granularity vertical integration and its impact on the micro-architecture design, where individual blocks are placed across multiple layers. By exploring multi layer implementations, our physical engine places blocks effectively within the given layer constraints, resulting in performance improvement and reduced power dissipation. Furthermore we make the following contributions:

- **Truly 3D Architectural Blocks:** We model components in multiple silicon layers and analyze the effects of different partitioning approaches on various components with respect to area, timing, power and temperature.
- **Architecture Driven Cube Packing Engine:** During the packing optimization, we introduce the process to choose the best configurations among a wide range of implementations.
- **3D Microarchitectural and Physical Design Co-optimization:** We extend the co-design framework[4, 32] to handle fine-grain 3D exploration. Given a frequency target, architectural netlist, and a pool of alternative block implementations, this framework finds the best solution in terms of performance (in BIPS), temperature, or both. Our state-of-the-art, temperature simulator tool is capable of optimizing the entire design by choosing from various alternative implementations of each block. We also perform automated thermal via insertion to help mitigate the impact of increased temperatures in 3D integration.

The rest of the paper is organized as follows: Section 2 briefly presents the background on three-dimensional integration technology, section 3 discusses alternative 3D implementations of components; the physical exploration approach along with cube packing algorithms are presented in section 4; section 5 provides experimental methodology and experimental results. Finally concluding remarks and future work are presented in section 6.

## 2. 3D IC Technology Background

3D IC fabrication refers to a wide range of technologies including multi-chip module (MCM) packaging [15, 16], chip-to-chip or wafer bonding [2, 3, 10], solid-phase re-crystallization [2], which have diverse characteristics in terms of circuit performance, manufacturing cost and thermal profile. There has been an extensive amount of work on 3D integration; hence we will only focus on closely related studies – [2] and [22] provide extensive discussions on current 3D research studies. In this work we focus on wafer bonding 3D IC technology; however the analysis and results can be extended to similar technologies. 3D integration commonly involves Face-to-Face (F2F) and Face-to-Back (F2B) approaches as illustrated in Figure 1. In case (a) the top device layer is flipped upside down so that the device layers are facing each other in F2F, and in case (b) the top device layer is oriented in the same direction as the bottom device layer in a F2B manner. We assume use of F2B, as shown in case (b).

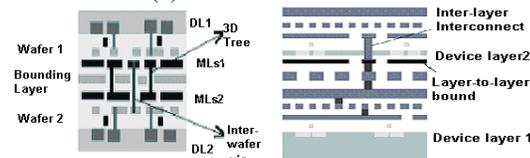

**Fig. 1 3D IC example with two device layers**
**(a) Face-to-Face approach (b) Face-to-Back approach**

## 3. Design of 3D Architecture Blocks

In this study we explore 3D microarchitecture design of modern complex out-of-order superscalar processors. Traditional design space explored by architecture-level design frequently involves different component sizes and characteristics along with various ways of connecting the components. This design space is extended by vertical integration technology to include aspects such as multi-layer implementations for each component and increased connectivity. We apply different partitioning techniques to model critical components in multiple layers and analyze their area, delay, and power consumption.

### 3.1. 3D Block Implementation Alternatives

We propose two main strategies for designing blocks in multiple silicon layers, in order to reduce intra-block interconnect latency and power consumption: *Block folding (BF)* and *Port partitioning (PP)*. *Block folding* implies folding of the block in X or Y direction – potentially shortening the wirelength in one direction, whereas *port partitioning* places the access ports of a cache-like structure in different layers.

As an example, we briefly describe the use of these strategies for cache-like blocks in our design driver architecture. In a typical cache-like structure, let us consider the case where each port contains bit, bitbar lines, a wordline, and two transistors per-bit. The wire pitch is five times the feature size [12,13] in most designs. For each extra port, the wirelength in both X and Y directions is increased by twice the wire pitch. On the other hand, the storage, which consists of 4 transistors, is twice the wire pitch in height, and has a width equal to the wire pitch. Hence, in general increased number of ports in a cache-like structure corresponds to larger port silicon area. Fig. 2(a) demonstrates a high-level view of a number of cache tag and data arrays connected via address and data buses. We make use of CACTI [12] to explore the design space of different subdivisions and find an optimal point for performance, power, and area.

**Block Folding(BF):** We consider two options for block folding: wordline folding and bitline folding. In the former approach the wordlines in a cache sub-array are divided and placed onto different silicon layers. The wordline driver is also duplicated. The gain from wordline folding comes from the shortened routing distance from pre-decoder to decoder and from output drivers to the edge of the cache. Similarly, bitline folding places bitlines into different layers but needs to duplicate the pass transistors. Our analysis shows that wordline folding has favorable access time and power dissipation in most cases compared to a realistic implementation of bitline folding. Hence in the following sections we only present results using wordline folding.

**Port Partitioning(PP):** Partitioning the ports and placing them on different layers provides advantages as shown in Fig. 3(c). Port partitioning allows reductions in both vertical and horizontal wire lengths. The width and height are simultaneously reduced by a factor of two, and the area by a factor of four. This reduces the total wire length and capacitance, which translates into a savings in access time and power consumption. Port partitioning requires vias to connect the memory cell to ports in other layers. In this design, 0.7um x 0.7um is allocated for each via. Via capacitance and resistance models are identical to [9].

Each block has different implementation alternatives, according to the number of layers and different partitioning strategy. We can evaluate various alternatives by our block modeling approach.

### 3.2 Block Modeling

3D-CACTI [9] is proposed as tool to enable three-dimensional exploration of caches and cache-like structures.

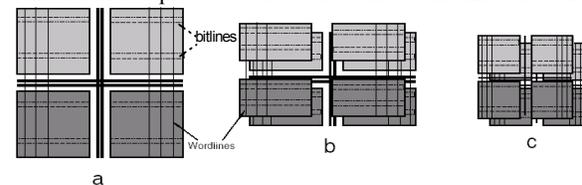

**Fig.2 3D block alternatives for a cache: (a) A 2D 2-ported cache; (b) Wordline Folding: Only Y-direction is reduced; (c) Port Partitioning: Ports are placed in two layers. Length in both X and Y is reduced.**

However, it is restricted to wordline/bitline folding. We extended the 3D-CACTI framework by adding the capability of exploring port partitioning and block folding. Furthermore, we added area estimation capability, including the area impact of 3D vias on the device layer. We validated our modifications to 3D-CACTI with HSpice.

We analyzed 3D implementations of various caches and cache-like structures including the instruction and data caches, register file(RF), load store unit, branch predictor and issue queue(IQ). Further details of each unit are presented in Section 5: Table 1. Among the experimented components, the issue queue has quite different characteristics compared to the rest of the cache-like structures. We used HSpice simulations with a model similar to Palacharla et al [11]. The area is approximated by 3D-CACTI using a similarly configured cache. For all the simulations the supply voltage is 1.0V and the technology size is 70nm. Transistor and wire scaling parameters are derived from [9, 17, 18], and we use copper interconnect in our simulations. Using these models, we quantified the gain of using 3D blocks in terms of area, timing, and power as shown in Fig. 3. The following points were observed during our experimental analysis:

- Area reduction: (Figure 3.a) Port partitioning is consistently more effective for area reduction over all structures. This is because port partitioning reduces lengths in both X and Y directions.
- Power and timing improvement (Figure 3.b-3.c): The power or timing improvement in port partitioning does not increase with the number of layers when the number of layers is larger than the number of ports. At the same time, the transistor layer must accommodate the size of vias. For Icache, Dcache and RF, 4-layer designs do not outperform 3-layer designs in terms of power or timing. On the other hand, with wordline folding, the trend continues with consistent improvement for more numbers of layers for most of the components. However, for IQs, the impact on match line wire length from stacking more layers increase the power consumption for folding to 9% with 4 layers.
- Block folding is more effective in reducing the block delay especially for the components with fewer ports. The data cache sees a 30% reduction in delay with BF, and a 23% reduction in delay with PP.

The diversity in benefit from these two approaches demonstrates the need for a tool to flexibly choose the appropriate implementation. The best 3D configuration of each component may not lead to the best 3D implementation for the whole system. Therefore, fine-grain 3D integration needs to be more than purely architectural design optimization or physical design optimization. To enable the co-optimization between 3D micro-architectural and physical design, we need a 3D co-design engine which can choose the implementation while executing the packing optimization.

## 4. Physical Exploration for 3D Micro-architecture

With the various implementations for each critical component, the architecture design is partially defined. Without the physical information, it is impossible to obtain the optimal implementations for components for the final chip. So that the co-optimization of architecture design and physical design should also be able to choose the configurations for components, such as the number of layers, the partitioning approach, etc. As described in the previous section, each component is not restricted to a rectangle, but it is likely to have cubic blocks, which have heights in Z-direction, in packing design. Therefore, cube packing algorithm should be developed to arrange the given rectangular boxes in a rectangular box of the minimum volume without overlapping each other. To evaluate a physical design for a processor, an efficient approach to estimate the system performance should be developed based on architecture structure and the model of components.

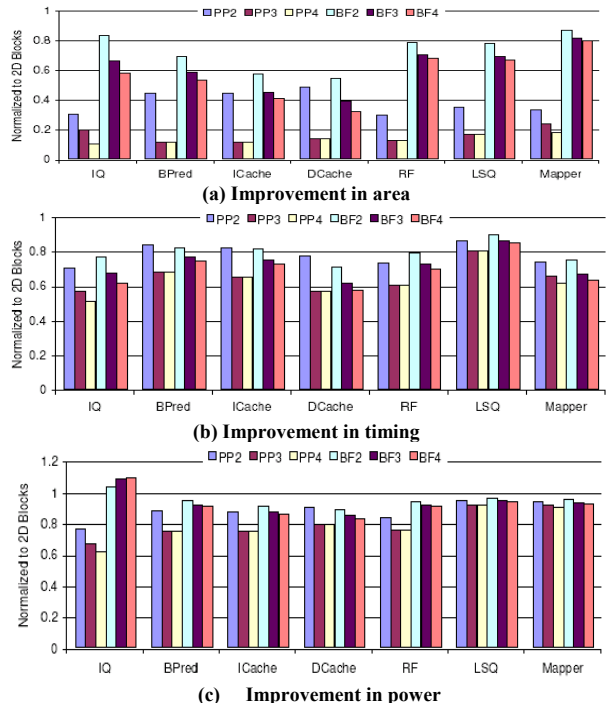

(a) Improvement in area

(b) Improvement in timing

(c) Improvement in power

**Fig.3 Improvements for multiplayer F2B design**

We propose an automated floorplanner that can be configured to optimize the packing for die area, performance, and temperature with consideration of interconnect pipelining. We extended CBL floorplanner based on simulated annealing scheme [23,25] as the cube packing engine, which is flexible enough to dynamically choose the configurations of blocks while doing the packing, and allows one to configure the number of device layers for 3D integration. Thermal via insertion and global routing can be employed to get the optimized thermal and routing profile. Once the exact block positions and wire latencies are known, this information is fed to our validation flow. A detailed cycle-accurate simulator that considers wire latency and power, and is coupled with power and thermal models, is used to validate the results from the MEVA-3D flow. In the sections that follow, we explain the components of our flow.

### 4.1 Performance Estimation

Our approach to calculate the performance of the processor during the floorplanning process is similar to the method used in [19]. As the target frequency during floorplanning is fixed, we want to calculate the IPC degradation caused by the extra latency introduced by the interconnects in the layout. The work by Borch et al. [20] showed that the IPC of a microarchitecture depends on a set of critical processor loops, and extra latency along these loops can cause the IPC to degrade. For each critical path, we develop IPC performance sensitivity models similar to the work by Sprangle et.al. [21] which provides information about IPC degradation due to extra latency along the path. Recently, this kind of exploration is used for performance improvement by Black et.al. in [29]. However this study is limited to single-layer blocks. During floorplanning, we calculate the total latency of each critical path including the blocks and the wires, and determine the total number of cycles at the target frequency required to cover this path latency. Extra latency from the wires is used to compute the new IPC, and hence the performance of the processor for that floorplan.

### 4.2 Cube Packing Engine

To enable the packing of 3D components, which may occupy more than one layer, we constructed an architecture-driven packing engine which is a true 3D packing engine The dimension in the Z direction represents the layer information. The 3D packing algorithm is extended from the CBL floorplanner[23][25].

### 4.2.1 Floorplanning for 3D micro-architecture

The floorplanning problem that we investigate here considers several components in its objective function that are important tradeoffs in 3D architectures. Specifically, we consider the die area (footprint), the performance of the microarchitecture in *BIPS*, the maximum on-chip temperature, and the wirelength so that the power from the interconnects can be reduced. Formally, we define the problem as follows:

**Given:**
(1) target cycle time $T_{cycle}$
(2) clocking overhead $T_{overhead}$
(3) target layer number of the chip $Z_{con}$
(4) list of blocks in the microarchitecture. Suppose for block $i$, there are $k$ different implementations which are recorded in candidate list as $\{c^i_1, c^i_2, ... c^i_k\}$. And each candidate $c^i_j$ has the width($w^i_j$), height($h^i_j$), layer number($z^i_j$), delay($d^i_j$) and power($p^i_j$).
(5) set of critical microarchitectural paths with performance sensitivity models for the paths

**Objective:** Generate a floorplan which optimizes for the die area, performance, and maximum on-chip temperature.

Fig.4 shows the optimization flow based on simulated annealing approach. Different with the previous floorplanning approach, we integrate the dynamically choosing of the blocks'configurations while doing the packing. Our cost function uses a weighted combination of area, performance, and temperature, and can be represented by

$$Cost = w1*\frac{1}{BIPS} + w2*Area + w3*Temp + w4*Wire$$

where BIPS corresponds to the performance of the microarchitecture with that floorplan of the blocks, Area is the total area of the floorplan. The performance (BIPS) is calculated through the method presented in previous section.

Temp corresponds to the maximum on-chip temperature based on CFD ACE+ temperature simulator [24]. The coefficients of w1, w2, w3 and w4 are used to control the different weight for each component. In our test evaluation, the performance component is given a high weight and will be optimized when the simulated annealing engine tries to minimize the cost function.

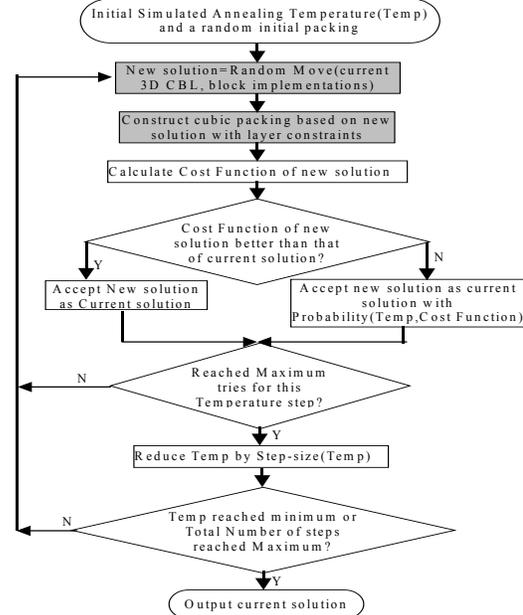

**Fig.4 Cube floorplanning flow based on 3D CBL**

### 4.2.2 3D CBL Representation

The topology of cube packing is a system of relative relations between pairs of 3D blocks in such a way as: block *A* is said to be left of block *B* when every point of *A* is left of every point of *B*. Relations of "right of", "above", "below", "front of" and "rear of" are analogously defined. 3D CBL uses three lists (*S,L,T*) to represent the topological relations between cubic blocks in 3D mosaic floorplan [25]. The 3D floorplan divides the total packing region into cubic rooms with sides. Each cubic room is assigned to no more than one cubic block. And rooms cover each other in X, Y or Z direction. Since we are representing the topological relations between blocks, the representation is independent of the sizes of blocks. Therefore, in our representation, if block *A* covers block *B*, block *A* is totally covered by a side and the side extension of block *B*.

Given a 3D mosaic floorplan, block *B* located at the upper-right-rear corner is defined as the ***corner cubic block*** since there is no other cubic block located at the right of, nor above nor behind this cubic block *B*. And corner cubic block *B* covers its neighboring blocks from the X, Y or Z direction. When the corner cubic block is packed in a certain direction, there are a series of blocks which are not yet covered by other previous blocks so that it can be covered by this cubic block. Therefore, we define the uncovered block list in packing sequence for each direction, which records the current available blocks to be covered. In Fig.5(b), before block 4 is inserted, the uncovered block list in Z-direction is

{1,2,3}, the uncovered block list in Y-direction is {1,3} and the uncovered block list in X-direction is {2,3}. Therefore, if the uncovered block list in a direction in packing sequence is {$B_1, B_2, \ldots B_k$}. Block $B$ is placed to cover $B_i$, then $B$ covers $B_m (m \geq i$: the block after $B_i$) and block $B$ does not cover $B_n (n < i$: the block before $B_i$). The uncovered block list can be updated dynamically with the process of corner cubic block. Since the blocks after block $B_i$ are covered by block $B$, they are no longer available to be covered in this direction. Hence, the updated uncovered block list should be {$B_1, B_2, \ldots B_{i-1}, B$}.

The information related to the packing process of the corner cubic block $B$ should include the following: the block's name, the orientation and the number of blocks covered by $B$ in the uncovered block list. To favor the generation of new solutions during the optimization process, we use a binary sequence $T_i$ to record the number of blocks covered by cubic block $B_i$. The number of 1s corresponds to the number of covered blocks. Each string of 1s is ended with a 0 to separate from the record of the next cubic block. Given a 3D packing, we delete the corner cubic block one by one and get the topological relations in the packing. At the end of deletion iterations, we have a sequence S of block names, a list L of orientations, and a list {$T_2, T_3, \ldots, T_n$} of covering information. The three-element triple $(S,L,T)$ composes a 3D CBL (as shown in Fig.5(c)).

To construct a floorplan based on given CBL, the blocks are packed from the left bottom-front corner to the upper-right-rear corner. For each insertion of corner block, the corner block is packed at the upper-right-rear corner of the current packing according to the corresponding direction in 3D CBL, and all the blocks packed before are at the left of or below or front of the current corner cubic block. In Fig.5(c), the corresponding 3D CBL list is given. Note that in a 3D CBL there are special cases where the number of successive "1" in list $T_i$ is greater than the number of the available uncovered blocks in the corresponding direction. To amend this, we automatically insert a "0" when the number of successive "1" in list $T$ is greater than the uncovered blocks, so that the block will cover all the available uncovered blocks. And we can construct the floorplan accordingly, based on an arbitrary 3D CBL list.

**Algorithm 3D CBL_Packing**
  Initialize the packing with cubic block $S[1]$;
  Initialize the uncovered lists in three directions;
  T-pointer = 0;
  **For** $i = 2$ to $n$:
    Uncovered_list = the uncovered list in $L[i-1]$ direction;
    $k$= the length of *Uncovered_list*;
    **While** $T[T\_pointer]$ ==1 and $k>0$:
      $S[i]$ coveres kth block and all blocks after kth block in uncovered_list from $L[i]$ direction and the coordinates of $S[i]$ is updated accordingly;
      $T\_pointer$++;
      $k$--;
  Update uncovered lists in all three directions: The blocks covered by $S[i]$ is deleted in uncovered list in $L[i-1]$ direction, $S[i]$ is added to all three uncovered lists.
**End.**

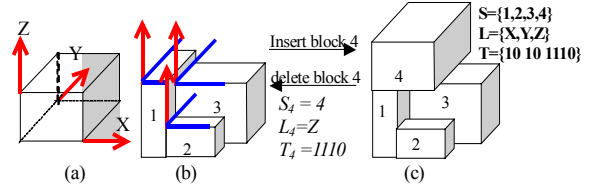

Fig.5 The process of corner cubic block: (a)X,Y,Z directions; (b) corner cubic block is 3 and uncovered block list in Z-direction is {1,2,3}; (c) corner cubic block is 4, uncovered block list in Z-direction is {4}.

Based on given 3D CBL list, we can construct the cubic floorplan accordingly in $O(n)$ time. But the complexities of ST [7] and 3D subTCG [8] are $O(n^3)$ in the worst case. Therefore, 3D CBL can get good results in a much shorter running time. And the major advantage of CBL representation is that the transformation from lists to packing is incrementally processed from lower left to upper right in linear time. Compared with graph-based representation, it is much easier to handle constraints by fixing the violations dynamically.

### 4.2.3 Packing optimization considering multiple 3D block implementation candidates

With 3D modeling in the previous section, the candidates vary in dimensions, delay, power consumption and layer numbers according to different partitioning approaches. In 3D microarchitecture design, the number of chip layers is often given as a constraint. Our approach adopts a standard simulated annealing process. Therefore, to choose the best feasible configuration for blocks, we define the new operation "*alternative_selection*" to create a new solution.
  **Operation: alternative_Selection:**
    Randomly choose a block *i* with multiple candidates;
    Randomly choose a feasible candidate from candidate list;
    Update block *i* with the dimension of the chosen candidate.

The move used to generate a neighboring solution is based on any one of the following operations:
1) Randomly exchange the order of the blocks in *S*;
2) Randomly choose a position in *L* and change the orientation;
3) Randomly choose a position in *T*, change "1" to "0" or change "0" to "1";
4) Alternative_Selection.

The various candidates of components enlarge the solution space by a great deal. Especially with some layer number constraints, parts of the solutions are infeasible. Therefore, heuristic methods are devised to speed up the searching process.

### 4.2.4 Packing with layer number constraints

During the packing process, the stacked blocks may violate the layer number constraints. The traditional method is to penalize the violations in cost function. However this method does not guarantee the feasibility of the final results and may slow down the convergence of the optimization. With 3D CBL representation, we pack the blocks in sequence. Therefore, we can dynamically change the blocks or CBL list during the packing. If some block exceeds the layer number constraint, we can fix the violation either by lowering the block or changing the direction of the block.

Given the number of layers of the design $Z_{con}$, we scan the CBL list to pack the blocks from lower-left-front corner to upper-right-rear corner. The coordinates of the lower-left-front corner for packed block B is $(x_B, y_B, z_B)$ with the corresponding implementation $c^B_j$. Hence the process can be described as following:

**Algorithm Fix_Violation**
**Input:**
B exceeding the layer number constraint: $z_B + z^B_j > Z_{con}$;
3D_CBL and the candidate list for block B.
**Output:** New 3D_CBL with current candidate selection $c^B$;
If $z_B < Z_{con}$
   For candidate $c^B_j$ in candidate list of B
     If $z_B + z^B_j \leq Z_{con}$
       choose this candidate $c^B = c^B_j$ and update the positions of B;
     return;
choose the candidate with the lowest Z-height and update the information of B;
If $L_B = Z$ // cover previous block from Z-direction
   Change $L_B$ to X or Y;
While $(z_B + z^B_j > Z_{con})$
   Increase the number of "1" in $T^{LB}_B$ which means the number of blocks covered by B in the direction $L_B$ is increased.
   Update the position of B;
End.

The extreme case is that block B is moved to the bottom ($z_B=0$). The candidate list should be constructed with the constraints that all the block's Z-height should be less than $Z_{con}$. Block B will not exceed the layer number constraint if $z_B=0$. Therefore, our algorithm will guarantee the feasibility of the results.

### 4.3 Performance Validation

Once we finished the physical planning stage, we can input the critical loop latencies and cycle time, along with the architectural configuration, into our cycle-accurate simulation framework. We adapted the SimpleScalar 3.0 tool set [26], a suite of functional and timing simulation tools for the Alpha AXP ISA, for our simulation framework. Our framework gives performance statistics in instructions per cycle (IPC) that can be combined with the cycle time from the floorplanning stage to give a result in BIPS.

## 5. Exploration with a Design Driver

We present the detailed evaluation results obtained for our design driver microarchitecture. Table 1 shows the baseline processor parameters used in this study. We modified SimpleScalar [26] to model this architecture. Based on [17], we assume that the clock cycle overhead is 46ps, which corresponds to roughly 1.8FO4 (fan-out-of-four) for 70nm technology. Thus, for a 4GHz target cycle time, we set the useful time for computation as 204ps and use this to calculate the number of pipeline stages required to cover a given path delay. The delay of interconnects is derived using the IPEM models [20] which consider several optimizations such as wire sizing, buffer insertion and buffer sizing, etc. To facilitate the insertion of repeaters, flip-flops, vias, etc., we assume that 10% of each block's area is reserved around the block in the floorplan. To perform our evaluation, results were collected for the SPEC2000 benchmarks.

**Table 1. Architectural parameters for the design driver**

| | |
|---|---|
| Processor Width | 6-way out-of-order superscalar, two integer execution clusters |
| Register Files | 128 entry integer(two replicated files), 128 entry FP |
| Data Cache | 8 KB 4-way set associative, 64B blocksize |
| Instruction Cache | 128KB 2-way set associative, 32B blocksize |
| L2 Cache | 4 banks, each 128KB 8-way set associative, 128B blocksize |
| Branch Predictor | 8K entry gshare and a 1K entry, 4-way BTB |
| Functional Units | 2 IntALU+1 IntMULT/DIV in each of two clusters; 1 FPALU and 1MULT/DIV |

### 5.1 Cube Packing Results

As described in the previous sections, we model each critical component with different implementations. Given the layer number constraints, our packing engine can pack the blocks successfully and choose the best implementation for each. In Fig.6, we show the packing results for 4GHz frequency. Fig.6(a) displays the best floorplan in terms of performance we achieved for one layer packing. The chip area is 4.9X4.9mm$^2$ and BIPS is 2.34. The runtime of the floorplanner is 344seconds. Fig.6 (b) display 3D view of the floorplan with the highest performance for 2 layers packing with 3D blocks. The area is 3.6x3.6mm$^2$. The runtime of the floorplanner is 3481seconds, in which most of runtime is spent on thermal evaluation. Our packing engine selects between single-layer or 2-layer block architectures. For blocks such as ALU, MUL and L2 cache units, single-layer implementation was selected. The rest of the blocks were implemented in 2-layer (We use cubic blocks to represent multi-layer block. All these multi-layer blocks are placed on multiple layers). A subset of blocks are partitioned by block folding and the remaining are port partitioning. By choosing the multi-layer components, the delay along the critical path can be reduced, and this leads to a better performance result. Table 2 shows the number of cycles along critical loops for different designs with 4GHz. Comparing the critical paths in Fig.6(a) and (b), the number of cycles along the branch misprediction loop is reduced from 21 to 15.

### 5.2 Performance Impact of 3D Integration

To study the impact of multi-layer blocks on the performance of the microarchitecture, we generated the best performance results for 2D block packing and 3D block packing by running the floorplanning engine 10 times and picking the best solution for each case. 2D blocks are restricted to a single layer of silicon, whereas the 3D architectural blocks span more than one layer of silicon using the wordline or port folding techniques. Fig.7 presents performance results relative to a single layer design driver. All three configurations (single layer, dual layer 2D blocks, dual layer 3D blocks) are running at 4GHz. On average, the use of 2D blocks in a 2-layer design improves performance by 6%. Since the blocks themselves do not take advantage of vertical integration, any performance gain can only come from a reduction in the inter-block wire latency. However, the overall reduction in path delay is not enough to reduce the loop by a cycle of our 4GHz clock. When we allow the selection of 3D block alternatives, we see a performance improvement of 23% on average over the single layer blocks to reduce the intra-block latency of critical processor loops as shown in Table 2. This result implies that, in this 4GHz case, using multi-layer blocks can further improve performance by about 16% over the case of using single-layer blocks alone, due to further reduction of intra-block latency. To explore the effect on the designs with different frequencies and layer numbers, in Table 3 we demonstrate the performance in BIPS when using different frequencies: 3GHz-6GHz and when using more silicon layers: 1 to 4 layers. Vertical integration with single-layer blocks can improve the performance about 19%. But if we allow the use of multi-layer blocks and optimize the implementation with the packing process, we can achieve a 36% performance improvement on average. In order to evaluate the sensitivity of our approach to different frequencies, we compile results in BIPS for the designs with multi-layer blocks in Fig.8. We can see that performance is getting better with the increase of the frequency and the number of layers. But when the frequency increases to 6GHz, the BIPS drops a little. That is because the higher the frequency of the chip, the more degradation the extra latency will have on chip performance. This trend is also true for single layer design and 3D design with single layer blocks.

**Table 2. Number of cycles along critical loops for 4GHz frequency:** 2D1L means 2D architectural blocks packing on one layer and 3D2L means 3D architectural blocks packing on two layers.

| | 2D1L | 2D2L | 3D2L | 3D3L | 3D4L |
|---|---|---|---|---|---|
| Wakeup | 5 | 4 | 4 | 3 | 3 |
| DL1 | 6 | 5 | 4 | 4 | 4 |
| L2 | 12 | 11 | 10 | 10 | 10 |
| Branch Misprediction | 21 | 18 | 15 | 16 | 14 |

Table 3. The performance comparison of 2D blocks and 3D blocks in BIPS for 3-6 GHz and 1-4 layer number

| Type of blocks | 1L | 2L 2D | 2L 3D | 3L 2D | 3L 3D | 4L 2D | 4L 3D |
|---|---|---|---|---|---|---|---|
| 3G | 2.09 | 2.2 | 2.70 | 2.38 | 2.83 | 2.8 | 2.91 |
| 4G | 2.34 | 2.48 | 2.91 | 2.76 | 3.05 | 2.83 | 3.25 |
| 5G | 2.48 | 2.65 | 3.19 | 3.01 | 3.40 | 3.2 | 3.58 |
| 6G | 2.34 | 2.53 | 3.16 | 2.96 | 3.33 | 3.29 | 3.52 |
| Compare | 1 | 1.07 | 1.29 | 1.20 | 1.36 | 1.31 | 1.43 |

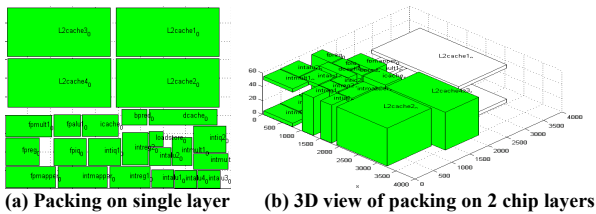

(a) Packing on single layer   (b) 3D view of packing on 2 chip layers

Fig.6 Cubic packing with different layers

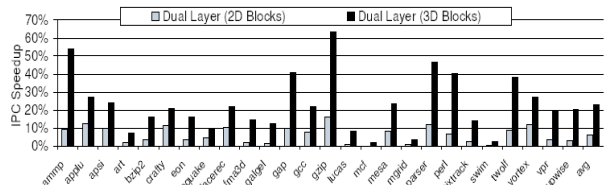

Fig.7 Performance speedup on SPEC2000 benchmarks

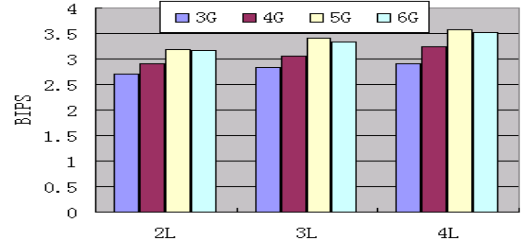

Fig.8 Frequency impact on performance in multi-layer implementations

### 5.3 Temperature Impact of 3D Integration

One of the major challenges of 3D integration is the increased thermal profile. Therefore, an accurate and fast thermal simulation framework is very crucial for design optimization. We use the finite element method (FEM) based CFD ACE+ temperature simulator [24] along with thermal via insertion [27]. Fig.9 illustrates the temperature comparison of the 2D and 3D architectural block technologies. The x-axis shows the different configurations with 2-4 silicon layers in the 3-5 GHz frequency range. The y-axis has the temperature in $^oC$ for 3D and 2D block technologies and the results of thermal via insertion. The ambient temperature is assumed to be 27 $^oC$. On average, multi-layer (3D) block configurations have 11% lower temperature.

Previous section shows that multi-layer blocks can save about 10-30% power consumption over single layer blocks. But temperature heavily relies on the layout. To relieve the hotspots, it is often necessary to keep potential hotspots away from one another. Even though single layer blocks may seem to have advantages over multi-layer blocks in this, our packing engine overcomes this issue by is intelligent layer selection for blocks depending on their thermal profile. Therefore, we can see that for 2-layer and 3-layer designs, the temperatures can be reduced due to the power reduction of multi-layer blocks and alternative selection in our engine.

Though multi-layer blocks can reduce some power consumption inside blocks, the temperatures still display a non-linear increase with an increased number of layers, as well as an elevation with higher frequencies. We see that without thermal via insertion, the temperatures are above $250^oC$ for 4 layers designs which are out of the normal operation range of silicon. [27] demonstrates the effects of thermal via insertion with floorplanning benchmarks: a 4-layer design with the peak temperature above 200 $^oC$ can be cooled to 77 $^oC$ using thermal vias. In our test, through effective use of thermal via insertion, the temperatures are reduced to around $100^oC$. Averagely, the thermal via insertion can lower peak temperature by about 60%. Therefore, by incorporating temperature-aware design planning, 3D architectures with multi-layer blocks provide 36% improvement in performance over 2D and 14% improvement over single-layer block 3D.

## 6. Conclusions and Future Work

Vertical integration has been shown to enable reduction both inter-block and intra-block wire latency. However, current research is limited to only exploiting inter-block latency due to lack of tool infrastructure. In this study we investigate the effects of using multi-layer blocks instead of constraining blocks in single layer silicon. Our results indicate that the effective use of multi-layer architectural blocks reduces the impact of wires within a block, through a reduction in block access time and/or power. On average we observed a 36% increase performance in BIPS compared to the single-layer case. Multi-layer block integration provides 14% improvement compared to the single-layer block case, along with 11% reduction in average temperature. Temperature-aware design planning and thermal vias enable on-chip temperatures of below 100$^o$C for 2-layer case. Our future work will consider the additional performance gain when using the timing slack from 3D integration to grow the sizes of architectural structures or make use of more power efficient but slower block alternatives.

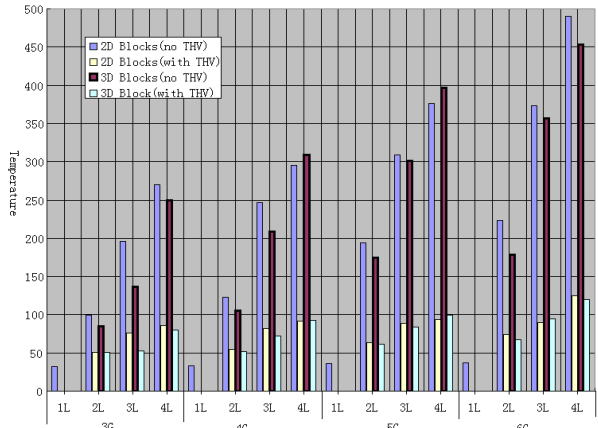

**Fig.9: Temperature comparison of 2D and 3D using thermal simulation for 3-5 GHz and 1-4 layer integration technology**